\documentclass[fleqn,usenatbib]{mnras}
\usepackage[T1]{fontenc}
\usepackage{ae,aecompl}
\usepackage{graphicx}   
\usepackage{amsmath}    
\usepackage{amssymb}    

\usepackage[landscape]{pdflscape}

\title[Physical parameters of AGN]{
Physical parameters of active galactic nuclei derived from properties of jet geometry transition region}

\author[Nokhrina et al.]{\parbox{\textwidth}{
E.~E.~Nokhrina$^{1}$\thanks{E-mail: nokhrina@phystech.edu},
Y.~Y.~Kovalev$^{2,1,3}$,
A.~B.~Pushkarev$^{4,2,1}$
}
\vspace{0.4cm}\\
\parbox{\textwidth}{
$^1$Moscow Institute of Physics and Technology, Dolgoprudny, Institutsky per., 9, Moscow region, 141700, Russia\\
$^2$Lebedev Physical Institute, Leninsky prosp.~53, Moscow, 119991, Russia\\
$^3$Max-Planck-Institut f\"ur Radioastronomie, Auf dem H\"ugel 69, 53121 Bonn, Germany\\
$^4$Crimean Astrophysical Observatory, Nauchny 298688, Crimea, Russia
}
}

\date{Accepted 2020 August 11; Received 2020 August 11; in original form 2020 May 28}

\pubyear{2020}

\begin{document}
\label{firstpage}
\pagerange{\pageref{firstpage}--\pageref{lastpage}}
\maketitle

\begin{abstract}
We use the observed jet boundary transition from parabolic to conical shape, which was earlier discovered as possibly a common effect in active galactic nuclei, to estimate a black hole, a jet and an ambient medium parameters. 
We explained earlier the geometry transition as a consequence of a change in the jet inner properties: a transition from a magnetically dominated to an equipartition regime. 
This interpretation allows us to estimate a black hole spin, a black hole mass and an ambient pressure amplitude, using the observed jet shape break position and the jet width at the transition point for 11 active galactic nuclei. The black hole spin values obtained using our method are consistent with the lower estimates for the sources with redshift $z<2$ from the spin evolution modelling. 
We find that the method of a black hole mass determination based on the relation between the broad-line region size and its luminosity may underestimate masses of the sources with large jet viewing angles. We propose a new method for the black hole mass determination, with the obtained masses being in interval $10^8-10^{10}\;M_{\odot}$. 
The range of the values of the ambient pressure amplitude points to the uniform medium conditions for the sources in our sample, with a tentative indication of higher pressure around FRII sources.    
\end{abstract}


\section{Introduction}
\label{s:intro}

Black hole and ambient medium properties play a crucial role in the models proposed to understand the jet activity of super massive black holes (SMBH) residing in the centres of active galactic nuclei (AGN). The direct measurements of these parameters are very difficult, and usually modelling is needed to derive estimates of these values. We propose to use observations of a jet boundary shape to estimate physical parameters using a semi-analytical modelling. In this paper we determine such parameters as a jet typical light cylinder radius, a black hole spin $a_*$ and ambient medium pressure through  semi-analytical modelling and measurements of a position of a jet shape change. We use the non-dimensional parameter $a_*=J/M^2$, where $J$ and $M$ are the black hole angular momentum and mass.

The most effective BH rotational energy extraction 
occurs when the rotational velocity of magnetic field lines $\Omega_{\rm F}$ equals to half of the black hole angular velocity $\Omega_{\rm H}$ \citep{BZ-77}. Numerical simulations were able to capture the Blandford--Znajek process as an excess of a jet power over an accretion power \citep{McKinney12} for extremely spinning black holes with the spin parameter $a_*>0.9$. On the other hand, simulations which model the SMBH mass and spin evolution predict extremely spinning BH at redshifts $z>2$ and moderate rotation down to $a_*\approx 0.1$ at lower redshifts \citep{Barausse12, Volonteri13, Sesana14}. \citet{Reynolds-13} and \citet{Brenneman-13} estimated some BH spins by modelling the $K\alpha$-line produced in a presumably cold thin accretion disc in the immediate vicinity of a BH produced by the reflection of the hot corona emission.
The method is applicable for the Seyfert I type AGNs, since it uses the broad line profile.
A less direct method based on the relation between a magnetic field, a total jet power and a BH spin was used by \citet{Daly-11, Daly-19} to calculate spins for a number of SMBH, including radio galaxies.

The role of the ambient medium in jet collimation evolved with the development of theory and numerical simulations. Indeed, the first elaborate models were focused on jet self-collimation \citep{BlP-82,HN89,PP92,LChB-92,ST-94} and, thus, disregarded the ambient medium, with jets being collimated by the hoop stress of a non-vanishing toroidal magnetic field due to the total electric current flowing in a jet. The numerical modelling accounts for the ambient pressure differently. \citet{Komissarov_etal09} 
employed solid walls to confine a relativistic outflow,
finding that the flows collimated to a parabolic shape 
can be accelerated effectively. There is a class of numerical models which include ambient pressure confining a jet while it propagates (drills) through the medium \citep{BTch-16, Nakamura+18} with possible formation of a cocoon and supplying it with mass \citep{Tchekhovskoy19}. In the latter work the resulting jet shape is parabolic on the scales up to $10^5$ gravitation radii in agreement with the observations of the M~87 jet \citep{Asada12}. The importance of the ambient pressure profile on the jet shape has been explored numerically by \citet{Tch-08,Komissarov_etal09,Lyu09}. 

A change in an observed jet boundary shape from wide to well collimated outflow on a scale of $10^2-10^3\;r_{\rm g}$ was reported by \citet{Junor_etal99} for M~87 and by \citet{RA_3C84} for 3C~84. \citet{Asada12} reported a jet shape transition from parabolic to conical on scales $10^5-10^6\;r_{\rm g}$ interpreting it as a result of a possible change in the external medium profile, supporting the importance of the external medium properties on the jet collimation \citep[e.g.][]{fromm_etal11, fromm_etal13}.
Currently, 13 nearby AGN are known to have such a jet shape break \citep{tseng16, Hada18, r:Akiyama18, r:Nakahara18, r:Nakahara19, r:Nakahara20}, including 10 sources by
\citet{Kovalev20_r1} identified using the stacked images technique \citep{MOJAVE_XIV} which revealed the geometry transition. 
The semi-analytical model with a total electric current confined inside a jet has been proposed by \citet{BCKN-17} to avoid having the current sheet at the jet boundary. This model effectively explains the transition from a quasi-parabolic to quasi-conical jet shape for the power-law ambient pressure profile, predicted by the Bondi accretion \citep{QN00, NF11}. Beskin model has been applied to explain the jet shape transition in M~87 \citep{nokhrina2019}, and we also use this approach here.

Using the interpretation proposed by \citet{Hada18, Kovalev20_r1}, we relate the transition of a jet geometry from parabolic to conical with a change in jet inner properties. In particular, within the model by \citet{BCKN-17}, we observe that the jet boundary changes its shape as the flow transits from a magnetically dominated to equipartition regime. This relation allows us to recover such AGN properties as a black hole spin, a black hole mass and ambient medium pressure. Assuming the absolute value of a SMBH spin to be in the expected range $\sim(0.1,\,0.99)$, we may also estimate the black hole mass with a precision of one order of magnitude.

This work is based on the source sample from \citet{Kovalev20_r1}. We use 12 sources with the detected jet shape geometry transition from parabolic to conical: two previously reported \citep{Asada12,Hada18}, and the other ten newly discovered by analysis of stacked images in \citet{Kovalev20_r1}. We omit in this study NGC 4162 \citep{r:Nakahara18} 
because of the way the authors used to assess the width of the outer jet. It is based on subtracting the low-brightness emission from the transverse profile which may lead to a progressive underestimation of the jet width at larger separations from the core. For the BL Lac object TXS 0815$-$094 the redshift is not known, so we work with 11 sources with a change in a jet shape. Their properties are listed in \autoref{t:break}. We also use additional 72 sources with a quasi-conical shape. Assuming that these sources might also have a jet shape geometry transition --- which has not been identified due to resolution constraints --- and that the position of this transition is bounded in the interval $(10^5-10^6)r_{\rm g}$ (gravitational radii) \citep{Kovalev20_r1}, we deduce some properties for them as well. 
The full list of the sources and their parameters 
is provided in Table~1 \citep{Kovalev20_r1}.

The structure of the paper is the following: in \autoref{s:breakpos}, we relate the measured jet shape geometry transition zone with the point calculated within our model to obtain the physical parameters of a system. We describe a sample of black hole masses in \autoref{s:BHM}. Using the mass values determined by different methods, we estimate the black hole spins in \autoref{ss:BHspins}. In \autoref{ss:mbhdiff}, we discuss the mass estimate method based on the relation between the broad-line region (BLR) size and luminosity, and  
we propose our method for constraining the values of black hole masses by measuring the jet shape transition position. We relate the latter to the jet total magnetic flux and ambient pressure amplitude in \autoref{ss:BreakPos}, and summarise our findings in \autoref{s:summary}.
Throughout this paper we will use the term ``core''  as the apparent origin of AGN jets which commonly appears as the brightest and most compact feature in VLBI images of blazars \citep[e.g.][]{Lobanov98_coreshift,Marscher08}. 
We adopt a cosmology with $\Omega_m=0.27$, $\Omega_\Lambda=0.73$ and $H_0=71$~km~s$^{-1}$~Mpc$^{-1}$ \citep{Komatsu09}.


\section{Geometry transition region}
\label{s:breakpos}

The detection of a jet boundary shape transition phenomenon in 13 nearby sources suggests that the transition of a jet shape from parabolic to conical might be a common effect, yet unresolved for more distant sources \citep{Kovalev20_r1}. The idea to connect a jet shape break with jet inner properties has been proposed by \citet{Kovalev20_r1} basing on the model by \citet{BCKN-17}. A jet shape is characterised by $k$-indices in dependence of the jet width $d$ on the distance $r$ along a jet: $d\propto r^{k}$. We use two indices for the sources with the clear detected shape transition from parabolic to conical: $k_1\approx 0.5$ in the quasi-parabolic domain and $k_2\approx 1.0$ in the quasi-conical domain. 

The model by \citet{BCKN-17} provides one of the possible solutions of a problem of an accurate transition between a jet outflow and an ambient medium. In order to omit having the current sheet at the jet boundary, \citet{BCKN-17} proposed to regard the warm flow with the following features: (i) the total electric current $I$ in a jet is closing due to a special choice of integrals conserved on magnetic surfaces $\Psi$, angular velocity $\Omega_{\rm F}(\Psi)$ and the angular momentum flux $L(\Psi)$; (ii) the integral of the energy flux $E(\Psi)$ has a thermal term, which is the only one left at the jet boundary defined by $\Psi=\Psi_0$. Such a model has several effects. Thermal effects are negligible up to the very boundary \citep{Kovalev20_r1}. The magnetic field vanishes at the jet boundary due to our choice of the integrals. The outer jet flow slows down to a full stop at the boundary. This may account for the observed slow sheath flow in the jets \citep{Mertens16}.

We assume that the jet boundary is at an equilibrium with an ambient medium: the pressure of an ambient medium is equal to the jet pressure. In general the latter is a combination of a magnetic field pressure, a hoop stress and a thermal pressure. In the cylindrical approach we neglect by the ram pressure. Within our model
the jet pressure at the boundary consists solely of thermal term balancing the external medium pressure. Close to the boundary the pressure balance holds \citet{BCKN-17}:
\begin{equation}
\frac{d}{dr_{\perp}}\left(\frac{B^2}{8\pi}+P\right)=0, \label{pressurebal}
\end{equation}
where the magnetic pressure is set by the total magnetic field amplitude $B$, the thermal pressure is $P$ and we use the cylindrical coordinates $\{r_{\perp},\,\varphi,\,r\}$. It is the conservation of a total pressure for a non-relativistic outer jet flow \citep{BCKN-17} allows us to relate the pressure balance at the jet boundary with a residual electric current inside a jet, since outside the light cylinder and up to the boundary a toroidal magnetic field dominates a poloidal field.
The major part of the total electric current in a jet is closed inside the bulk jet volume within this model. As only a small residual current is left close to the boundary \citep{BCKN-17}, \autoref{pressurebal} shows that the jet pressure at the boundary is smaller than previously anticipated. This fact allows us to reproduce the jet shapes with a predicted break for the pressure amplitudes close to the measured value around a jet in M87 \citep{RF15,nokhrina2019, Kovalev20_r1}. 

Within our model we can calculate the jet pressure at the boundary as a function of a jet width.
Calculated pressure as a function of a jet width $P(d)$ behaves as two power-laws with the transition region between them. \citet{Kovalev20_r1} showed that the transition between two power-laws coincides with the domain where the initially magnetized outflow reaches equipartition, with the Poynting flux being equal to the plasma bulk kinetic energy flux. We connect the this with a different behaviour of an electric current in magnetically-dominated and particle-dominated regimes. As the flow is accelerating, an electric current is closing due to a transformation of a Poynting flux (associated with a toroidal field and thus an electric current) to a particle bulk kinetic energy flux. The residual of this current defines a jet pressure at a boundary. As the effective acceleration ceases after a flow reaches a local magnetization a value of unity, a rate of an electric current closure changes. This impacts a pressure behaviour at the jet boundary.

For the ambient pressure corresponding to the 
Bondi accretion
\begin{equation}
P=P_0\left(\frac{r}{r_0}\right)^{-b}, \label{Pscale} 
\end{equation}
where $b\approx 2$ \citep[see discussion in][]{nokhrina2019}, the predicted jet boundary shapes are
$d\propto r^{0.5}$ upstream the break and $d\propto r^{0.9}$ downstream, which is in agreement with the observations of nearby sources \citep{Kovalev20_r1}. 

If our interpretation is correct and the break in a jet boundary shape corresponds to the transition of the outflow
from the magnetically dominated to equipartition regime, then the jet transverse radius at the jet geometry transition region (GTR) and its position may provide us information about the central engine, the jet and the outer medium properties. As shown before \citep[see e.g.][]{Beskin06, TMN09, Lyu09}, the ideal MHD numerical and semi-analytical models predict the linear growth of the flow Lorentz factor with the jet radius until
the Poynting flux becomes approximately equal to the particle kinetic energy flux.
The saturation of this ``ideal'' acceleration is reached at the jet radius 
\begin{equation}
{r_{\perp}}\approx \sigma_{\rm M}R_{\rm L},
\label{gamma_lin}
\end{equation} 
where the initial magnetization $\sigma_{\rm M}$ --- the ratio of Poynting flux to a particle kinetic energy flux at the jet base,
and the light cylinder radius is set by an angular velocity $R_{\rm L}=c/\Omega_{\rm F}(0)$.
The good agreement of the 
estimates on the Michel's magnetization parameter, obtained by \citet{NBKZ15}, with the implied typical flow Lorentz factors supports it.

\begin{table*}
\caption{Parameters of the sources with a jet shape transition, compiled from \citet{Hada18} for 0321$+$340 (1H 0323$+$342), \citet{nokhrina2019} for 1228$+$026 (M~87) and \citet{Kovalev20_r1} for the rest of the sources.
The columns are as follows: 
(1) source name (B1950);
(2) redshift \citep[collected by][]{Kovalev20_r1};
(3) viewing angle \citep[collected by][]{Kovalev20_r1}; 
(4) black hole mass estimated from kinematics (stellar, gas or cluster velocity dispersion) method; 
(5) black hole mass estimated basing on the assumption of virialised broad lines region movement and correlation between the size of BLR and UV/optical luminosity (for 0321$+$340 the second value), the fundamental plane method (2200$+$420) and relation between BH mass and buldge luminosity (0321$+$340, first value);
(6) jet width at the break; 
(7) deprojected distance of a break from a BH along the jet; (8) $k$-index upstream the jet shape geometry transition (parabolic flow);
(9) $k$-index downstream the jet shape geometry transition (conical flow); (10) Fanaroff--Riley (FR) class; (11) FR class reference: [1] \citet{ConBrod-88}, [2] \citet{CLK07}, [3] \citet{ABM-08}, [4] \citet{LP-84}, [5] \citet{WBU-87}, [6] \citet{CRATES}, [7] \citet{Owen_etal00}, [8] \citet{NVSS}, [9] \citet{DRAGN}, [10] \citet{Cassaro-99}, [11] \citet{Antonucci-86}.
\label{t:break}}
  \begin{tabular}{ccrcllrcccc}
  \hline\hline
 Source & z & $\theta_{\rm obs}$ & $M_{1}$ & $M_{2}$ & {$d_{\rm break}$} & $r_{\rm break}^{\rm deproj}$ & $k_1$ & $k_2$ & FR & FR \\
       & & (deg) & $\left(\log M_{\odot}\right)$ & $\left(\log M_{\odot}\right)$ & (pc)              & (pc)                         &       &    & class & reference \\
 (1)    & (2)               & (3)                   & (4)  & (5) & (6) & (7) & (8) & (9) & (10) & (11) \\ 
\hline
 0111$+$021 & 0.047 & 5.0 & \ldots & \ldots & $0.28\pm 0.03$ & $27.31$ & $0.495\pm 0.077$ & $0.934\pm 0.054$ & \ldots & [1] \\
 0238$-$084 & 0.005 & 49.0 & 8.19 & 5.51 & $0.05\pm 0.01$ & $0.49$  & $0.391\pm 0.048$ & $1.052\pm 0.081$ & I & [2] \\
 0321$+$340 & 0.061 & 6.3 & \ldots & 8.6 (7.30) & 1.16           & $106.07$ & $0.6$            & $1.41$      & I & [3] \\
 0415$+$379 & 0.049 & 13.4 & \ldots & 8.21 & $0.74\pm 0.03$ & $29.00$ & $0.468\pm 0.026$ & $1.175\pm 0.046$ & II & [4] \\
 0430$+$052 & 0.033 & 18.7 & 8.13 & 7.52 & $0.29\pm 0.04$ & $5.77$  & $0.556\pm 0.070$ & $1.131\pm 0.027$ & I & [5] \\
 1133$+$704 & 0.045 & 5.0 & 8.21 & \ldots & $0.50\pm 0.02$ & $14.80$ & $0.528\pm 0.040$ & $0.828\pm 0.047$ & \dots & [6] \\
1228$+$126 & 0.004 & 14.0 & 9.82 & \ldots & $1.24\pm 0.04$ & $43.00$  & $0.57$           & $0.90$      & I &  [7]  \\
 1514$+$004 & 0.052 & 15.0 & \ldots & \ldots & $0.34\pm 0.02$ & $13.10$ & $0.564\pm 0.048$ & $0.886\pm 0.022$ & \ldots & [8] \\
 1637$+$826 & 0.024 & 18.0 & 8.78 & \ldots & $0.16\pm 0.01$ & $3.30$  & $0.506\pm 0.041$ & $0.730\pm 0.029$ & I & [9] \\
 1807$+$698 & 0.051 & 7.3 & 8.51 & 7.14 & $0.25\pm 0.04$ & $12.83$ & $0.388\pm 0.087$ & $1.023\pm 0.025$ & II & [10] \\
 2200$+$420 & 0.069 & 7.6 & \ldots & 8.23 & $0.95\pm 0.04$ & $24.57$ & $0.537\pm 0.057$ & $1.124\pm 0.009$ & I & [11] \\
\hline
\end{tabular}
\end{table*}

We propose to estimate the central source and jet parameters within our model using
the observations of a change in a jet shape.
We calculate the jet boundary form for a set of values of the Michel's
magnetization parameter $\sigma_{\rm M}$. In non-dimensional units it is the only parameter which defines the solution for the given integrals. We calculate the profile $P(d)$ and fit it with two power laws, which intersect at the point with non-dimensional ``coordinates'': the predicted jet width, normalised by the light cylinder radius, $d_*(\sigma_{\rm M})$ and the predicted non-dimensional pressure $P_*(\sigma_{\rm M})$ at the GTR. Returning to the dimensional variables, we associate this model transition point with the observed one in the jet shape as follows:
the non-dimensional jet width at the break is related to the measured jet width $d_{\rm break}$ at the geometry transition region as
\begin{equation}
d_*(\sigma_{\rm M})=\frac{d_{\rm break}}{2R_{\rm L}}\label{d_star}
\end{equation}
and the non-dimensional outer pressure at the GTR relates to the jet pressure $P_{\rm break}$
at the boundary as
\begin{equation}
P_*(\sigma_{\rm M})=\frac{P_{\rm break}}{p_0},
\label{P_star}
\end{equation}
where the pressure is normalised by the value
\begin{equation}
p_0=\left(\frac{\Psi_0}{2\pi R_{\rm L}^2\sigma_{\rm M}}\right)^2, \label{Pbrdim} \end{equation}
with $\Psi_0$ being the total magnetic flux in a jet.
The magnetic field scale is set by the value $\Psi_0/(\sqrt{\pi/2} R_{\rm L}^2\sigma_{\rm M})$. 
From \autoref{Pscale} and \autoref{Pbrdim} the position of a break along the jet is given by
\begin{equation}
r_\mathrm{break}=r_0\left[\frac{P_*}{P_0}\left(\frac{\Psi_0}{2\pi R_{L}^2\sigma_{\rm M}}\right)^2\right]^{-1/b}. 
\label{rbreak}
\end{equation}
Calculating $d_*$ and $P_*$ for well-confined $\sigma_{\rm M}$,
we may reconstruct from \autoref{d_star} the light cylinder radius, and from \autoref{rbreak} the pressure magnitude $P_0$ at a given distance $r_0$.

The particular estimates of these parameters depend on a value of initial magnetization $\sigma_{\rm M}$. This parameter is equal to the maximum Lorentz factor $\gamma_{\rm max}=\sigma_{\rm M}$ attained by the bulk flow motion if all the electromagnetic energy is transformed into particle kinetic energy. Semi-analytical and numerical modelling show that
relativistic jets accelerate effectively only up to equipartition, which corresponds to $\gamma\sim\sigma_{\rm M}/2$. This argument can constrain the magnetization parameter for the sources with the detected superluminal motion. Unfortunately, this is not the case for six sources with the observed recollimation \autoref{tableParam}, where no apparent superluminal motion is detected.

Another way to constrain the magnetization is based on fitting the jet transverse causality parameter $\gamma\theta_{j}$ related to a jet half-opening angle $\theta_j$. For $\gamma\theta_{j}\lesssim 1$ the flow is causally connected across a jet, meaning that the disturbances at the jet boundary, moving with fast magneto-sound speed, can reach the jet axis. Modelling provides that
effectively accelerating jets satisfy this condition
\citep{TMN09,Komissarov_etal09}, with the Lorentz factor growing as $\gamma\propto r_{\perp}$. The observations yield the median value $\gamma\theta_j=0.17$ \citep{MOJAVE_XIV}, where ${\rm tan}\,\theta_j=d/2r$. In our modelling we can calculate this parameter along a jet. The resultant product $\gamma\theta_j$ assumes an approximately constant value in the conical
domain, where the acceleration ceases. Thus we may choose $\sigma_{\rm M}$ so that the asymptotic of $\gamma\theta_j$ tends to the value $0.17$. For the sources with a detected superluminal motion we check the choice of $\sigma_{\rm M}$ for consistency against an observed Lorentz factor.

\section{Black hole masses sample}
\label{s:BHM}

In order to relate the measured distances to the physical values, such as gravitational radius,
we need to estimate the masses of the supermassive black holes residing in our sources.
There are several methods for BH mass estimates  
\citep[see e.g.][]{WU02}. Two of the most reliable ones are kinematics (stellar, cluster or gas velocity dispersion) and the reverberation mapping. 
The latter employs velocities of broad-line region (BLR) clouds in the central BH gravitational potential, with the size of BLR determined by the measured time lag between the ionizing continuum and the broad-line strength. 
A less accurate method for the BLR size determination is based on the empirical relation between the broad-line region size and spectral luminosity (see \citet{WU02} and references therein). This method of mass determination uses the measurements of spectral luminosity $\lambda L_{\lambda}$ at wavelength $\lambda$ (in angstr\"oms) and measurements of full width of an optical line emission at half-maximum (FWHM) to determine the size of a broad-line region and a gas velocity. This method provides BH mass estimates for most of the sources in our sample. Below we designate as $M_1$ the SMBH mass deduced from the kinematics and $M_2$ from other methods described above.

To estimate the BH mass $M_{2}$ by relation between the BLR size and luminosity in lines, we use mainly the observations by \citet{Trrlb12} with few exceptions \citep[see references in the Table~1 in][]{Kovalev20_r1}.
We use the black hole mass estimators for H$\beta$ and MgII lines by \citet{McLJ02}: 
\begin{equation}
\frac{M_{\rm BH}}{M_{\odot}}=4.74\left(\frac{\lambda L_{5100}}{10^{44}\; {\rm erg\;s^{-1}}}\right)^{0.61}
\left(\frac{\rm FWHM_{H\beta}}{{\rm km\;s^{-1}}}\right)^{2}, 
\end{equation} 
\begin{equation}
\frac{M_{\rm BH}}{M_{\odot}}=3.37\left(\frac{\lambda L_{3000}}{10^{44}\; {\rm erg\;s^{-1}}}\right)^{0.47}
\left(\frac{\rm FWHM_{\rm MgII}}{{\rm km\;s^{-1}}}\right)^{2},
\end{equation} 
and for CIV line by \citet{VP06}:
\begin{equation}
\frac{M_{\rm BH}}{M_{\odot}}=4.57\left(\frac{\lambda L_{1350}}{10^{44}\; {\rm erg\;s^{-1}}}\right)^{0.53}
\left(\frac{\rm FWHM_{\rm CIV}}{{\rm km\;s^{-1}}}\right)^{2}.
\end{equation} 
For the sources with the information for several lines we used preferentially H$\beta$ measurements. We  checked that all such sources have a redshift
$z<0.9$ for a reliable optical line detection. There are 45 sources in our sample with a mass obtained by this method.

The only sources with the masses $M_1$ obtained from the 
direct stellar velocity dispersion measurements, stellar kinematics and gas kinematics are the sources with the confirmed jet shape break (see the 4th column in \autoref{t:break}). There are
six such sources in total (see \autoref{t:break}). We observe a clear dichotomy in mass values for the sources with available masses $M_1$ and $M_2$ obtained by different methods. The masses $M_1$ from the kinematics are greater than $M_2$ in our sample.

For the mass of a BH in 0321$+$340 (1H 0323$+$342) we use two different estimates.
The relation between the BLR velocities and optical line luminosity provides a low value 
$M=10^{7.3}M_{\odot}$ (the second value in the 5th column in \autoref{t:break}).
However, the analysis based on the population properties by \citet{LeonTavares14} and the consistency of a jet shape break position \citep{Hada18, Kovalev20_r1} point to a possible mass underestimate by this method.
So, we use also for 0321$+$340 the mass provided by \citet{LeonTavares14} basing on the scaling relations between the black hole mass and the luminosity of a spheroid: $M=10^{8.6}M_{\odot}$ (the first value in the 5th column in \autoref{t:break}).

For the sources with a detected jet shape transition we use all the available mass measurements. For the sources with the detected conical shape and the presumed transition from the parabolic form we will use the correlation between the BLR velocities and luminosity in optical lines to estimate their masses for the uniformity of the sample. 

\section{Black hole spins}
\label{ss:BHspins}

The jet width $d_{\rm break}$ at the break alone gives us directly the size of a light cylinder $R_{\rm L}$. It may also provide information about the absolute value of the BH
spin. Indeed, the initial magnetization is very well constrained by the jet kinematics \citep{MOJAVE_XVII} and the core-shift effect \citep{NBKZ15}, yielding $\sigma_{\rm M}<50$ for the majority of the AGNs. For a range $\sigma_{\rm M}\in[5;\;50]$, the non-dimensional model jet radius at the GTR $d_*$
changes by a factor of four, allowing us to estimate the light cylinder radius
\begin{equation}
R_{\rm L}=\frac{d_{\rm break}}{2d_*(\sigma_{\rm M})}\label{RL}
\end{equation}
for the observed $d_{
\rm break}$ with the same accuracy. This estimate may be improved if the initial magnetization can be better constrained by the arguments in  \autoref{s:breakpos}. 

The light cylinder radius $R_{\rm L}=c/\Omega_{\rm F}$ can be related to the gravitation radius $r_g=GM/c^2$ and the black hole spin 
using the relation $\Omega_{\rm F}=\Omega_{\rm H}/2$, with the proportionality coefficient $1/2$
for the condition of maximum power output of the Blandford--Znajek process \citep{BZ-77}. The BH spin parameter relates to $R_{\rm L}$ and $r_{g}$ as
\begin{equation}
a_*=\frac{8(r_g/R_{\rm L})}{1+16(r_g/R_{\rm L})^2}\label{astar}.
\end{equation}

The generally expected range of absolute spin values in active galaxies is $|a_*|\in(0.1;\;1)$. The numerical simulations predict
spins greater than $0.9$ for the Blandford--Znajek process to operate effectively \citep{McKinney12}. However,
the studies of BH mass and spin evolution due to accretion and mergers predict more moderate spins for the sources with a redshift $z<2$: $|a_*|\in(0.1,\,0.7)$, with a tendency to lower spins for higher-mass black holes \citep{Barausse12, Volonteri13, Sesana14}.  These evolution tracks do not account for a momentum taken away by the jet itself. Indeed, the rate of momentum carried by a jet is given by 
\begin{equation}
\frac{d}{dt}{\cal{L}}=\frac{1}{c}\int_0^{\Psi_0}L(\Psi)d\Psi,
\end{equation}
and the total electromagnetic energy losses by a jet at its base (which for high magnetization is a good estimate for the total jet power) are given by
\begin{equation}
W_j=\frac{1}{c}\int_0^{\Psi_0}\Omega_{\rm F}L(\Psi)d\Psi.
\end{equation}
The zero-order estimate provides
$\dot{\cal{L}}=2W_j/\Omega_{\rm H}$. For fiducial $a_*=0.1$, $M_{\rm BH}=10^9\,{\rm M}_{\odot}$ and $W_j=10^{43}\;{\rm erg\;s^{-1}}$ the momentum loss is of the order of $10^{48}\;{\rm g\;cm^{2}\;s^{-2}}$.
We would expect for our close sources the BH spin of the order of $0.1 - 0.7$. This value may be lower if a jet carries away the momentum effectively, although it depends on the accretion rate \citep{Moderski96}.

\begin{figure}
\centering
\includegraphics[width=\columnwidth,trim=0cm 0.8cm 0cm 0cm]{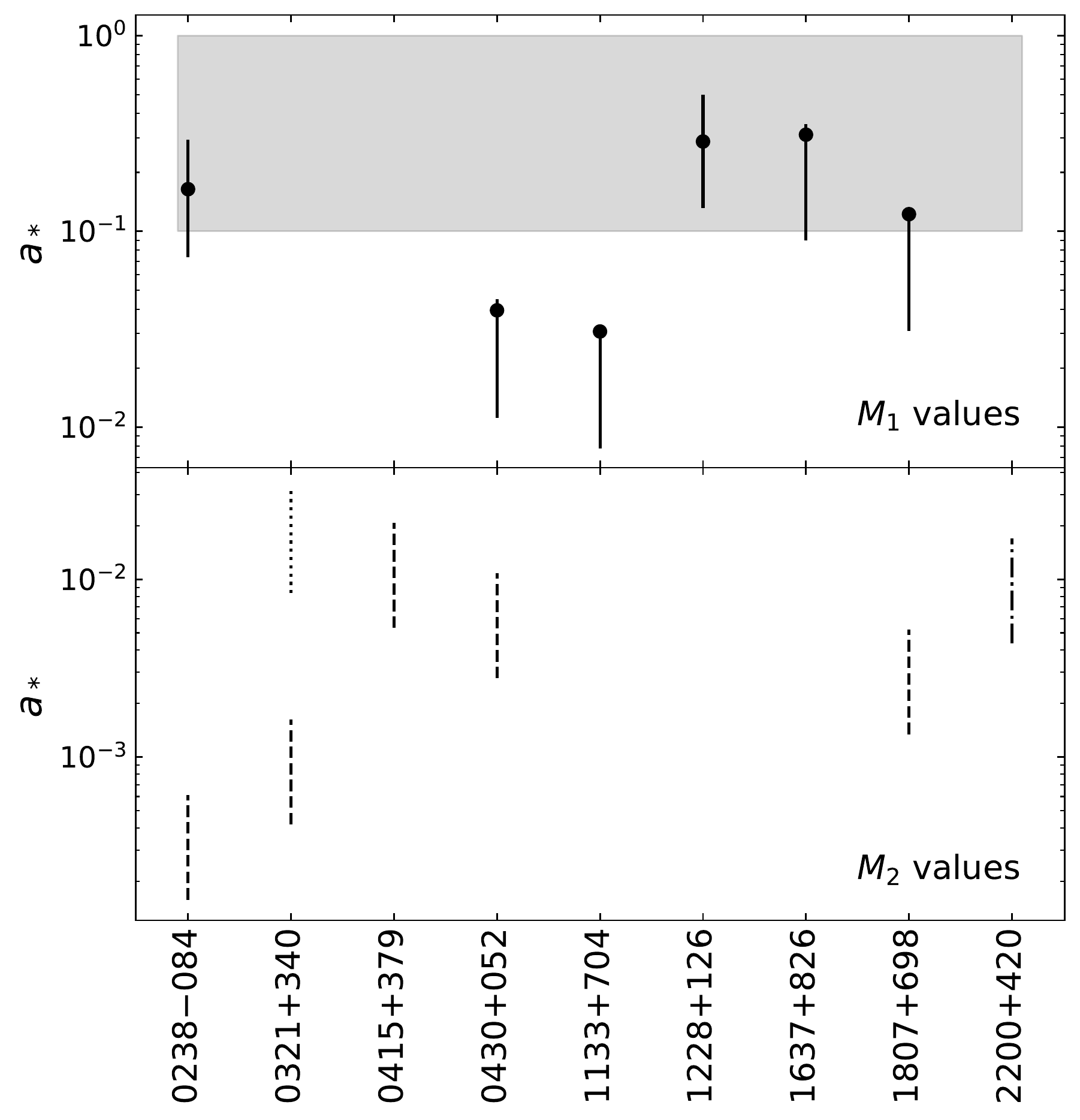}
\caption{Estimated spins for nine sources with the detected jet shape break and known masses. The grey band designates the spin range $a_*\in(0.1;\;0.99)$. The vertical lines are the spin range for each source, corresponding to the magnetization range from $\sigma_\mathrm{M}=5$ (lower end) to $\sigma_\mathrm{ M}=50$ (upper end). The upper panel corresponds to the sources with the mass $M_1$ (solid lines). The spin values estimated for the choice of $\sigma_{\rm M}$ in \autoref{tableParam} are highlighted by the filled circles. The lower panel --- for masses $M_2$ by other methods. The dotted line designate the mass obtained by the buldge luminosity; the dashed lines correspond to the mass by the relation between the BLR size and the line luminosity; the dashed--dotted line corresponds to the mass obtained by the fundamental plane method \citep[see details in][]{WU02}.}
\label{spins_break}
\end{figure}

For the sources with the measured jet width at the GTR we assume the black hole mass values collected in \autoref{t:break}, determined by the velocity dispersion method, the BLR size-luminosity relation method, the bulge luminosity and by the fundamental plane method.
We calculate the non-dimensional model jet radius at the break $d_*$ for different magnetizations. We have the measured jet width at the break $d_{\rm break}$ and the gravitational radius $r_{\rm g}$. Combining the model prediction with the observations and using equations (\ref{RL}) and (\ref{astar}) we can estimate the BH spin. In what follows, we plot the BH spin parameter in \autoref{spins_break} for a range $\sigma_{\rm M}\in[5;\;50]$. We observe that the spin range for four sources out of nine fit partially into the band $|a_*|\in(0.1;\;0.99)$ of the expected spin values. We also observe that the spin values based on the mass values obtained by the direct velocity dispersion method tend to be an order of magnitude higher than those based on the masses measured by other methods,
with the number of sources insufficient to make a definite conclusion on this point.

\begin{figure}
\centering
\includegraphics[width=\columnwidth, trim=0cm 1cm 0cm 0cm]{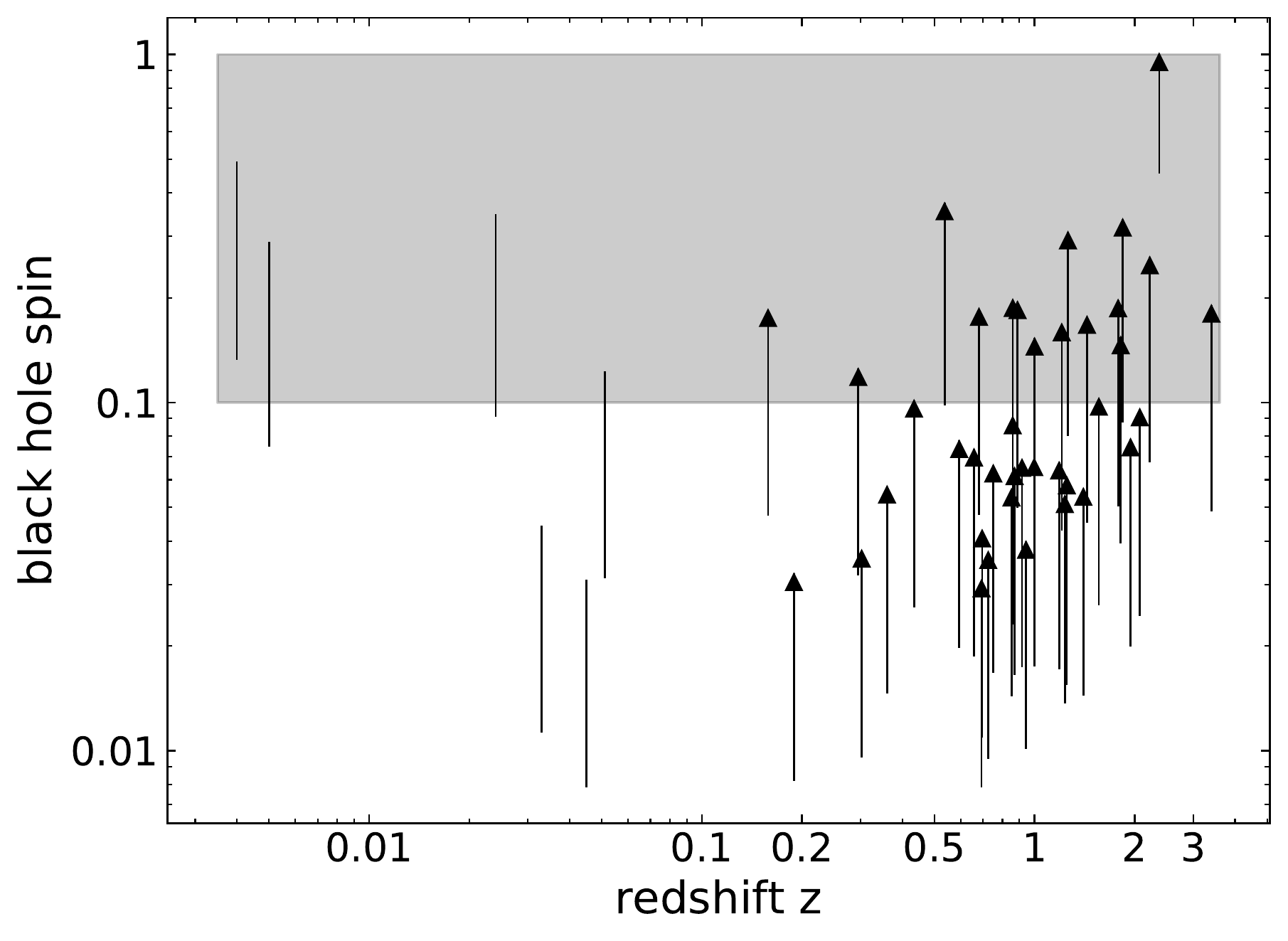}
\caption{Spin values for 39 sources with the masses $M_2$ as a function of redshift. The core width was used as an upper estimate for a jet width at the break. This means that the plotted spins are lower estimates which is designated by the arrow signs. The magnetization range is from $\sigma_{\rm M}=5$ (lower end) to $\sigma_{\rm M}=50$ (upper end). We also plot six spin values from the upper panel in \autoref{spins_break}.}
\label{spins_rm}
\end{figure}

The obtained spin range may be narrowed. Constraining the magnetization parameter for each source as either to reproduce the observed maximum Lorentz factor or to reproduce the causal parameter asymptotic behaviour $\gamma\theta_j\approx 0.17$ \citep{MOJAVE_XIV}, we constrain the individual spins for the sources.
In the domain of a jet boundary conical shape the acceleration saturates, so the parameter $\gamma\theta_j$ tends to the constant value. We choose $\sigma_{\rm M}$ to reproduce the value $\gamma\theta_j\approx 0.17$ and estimate the black hole spin absolute value $|a_*|$. We see that larger values of $\sigma_{\rm M}$ in a range 20$-$50 are preferred because they provide the values of the causality parameter $\gamma\theta_j$ close to the observed median. Thus the chosen values for $\sigma_{\rm M}$ are presented in the third column in \autoref{tableParam}, estimates on $|a_*|$ are in the seventh column and $\gamma\theta_j$ are in the nineth column, correspondingly.
We designate by the filled circle the spin value from \autoref{tableParam} for the sources in the upper panel of \autoref{spins_break}.
The mean for the spins from \autoref{tableParam} is 0.16, and the median is 0.14 for the sources with the mass $M_1$ obtained by the kinematics method. These values correspond to the lower spin value expected from the spin evolution models \citep{Barausse12, Volonteri13, Sesana14}.

\begin{table*}
\caption{Model and derived jet parameters for known masses of a central black hole.
The columns are as follows: (1) the source name (B1950); (2) the observed Lorentz factor, basing on $\beta_{\rm app}$ and $\theta_{\rm obs}$ from Table~1 \citep{Kovalev20_r1}; (3) the Michel's magnetization parameter, which we set to obtain either realistic $\gamma_{\rm max}$ or $\gamma\theta_{\rm j}$; (4) the light cylinder radius derived for the chosen $\sigma_{\rm M}$; (5) the total jet power obtained by the relation from \citep{Cav-10} except for M87, for which we estimate the total power using our model; (6) the reference for the flux density at 300$-$400 MHz: [1] \citet{TXS96}, [2] \citet{WENSS}, [3] \citet{DRAGN}, [4] \citet{Ku81r}; (7) the rotational parameter --- the ratio of the gravitational radius to the light cylinder radius; the first line --- using the mass obtained by the velocity dispersion method (including fundamental plane method for BL Lac); the second line --- by a reverberation method; (8) the black hole spin; the same as for the rotation parameter; (9) the maximum value for $\gamma\theta_{\rm j}$ along the jet, where $\theta_{\rm j}$ is the jet opening angle found by
semi-analytical modelling; (10) the total magnetic flux contained in the jet obtained using \autoref{Pjet}, except for M87; (11) the pressure $P_{\rm break}$ at the break point, the corresponding parameter for M87 is calculated for the parameters measured by \citet{RF15}; (12) the black hole mass estimate based on the jet width at the break (\autoref{ss:mbhdiff}).
\label{tableParam}}
\begin{tabular}{lrcccccccccc}
\hline\hline
Source & $\gamma_{\rm obs}$ & $\sigma_{\rm M}$ & $R_{\rm L}$ & $\log_{10}W_{\rm j}$ & $W_{\rm j}$ re- & $a$ & $a_*$ & $\gamma\theta_{\rm j}$ & $\log_{10}\Psi_{0}$ & $\log_{10}P_{\rm break}$ & $\log_{10}M$ \\
(B1950) &&               & ($10^{-3}$~pc)  & (erg/s) & ference &     &       & &  (${\rm G\,cm^2}$) & (${\rm dyn\,/\,cm^{2}}$) & $(M_{\odot})$ \\
 (1) & (2)              & (3)               & (4)             & (5) & (6)   & (7)                    & (8) & (9) & (10) & (11) & (12) \\
\hline
0111$+$021 & 1 & 50 & 1.18 & $44.57$ & [2] & \ldots & \ldots & 0.04 & 33.54 & $-5.26$ & 8.4--9.7\\
 0238$-$084 & 1 & {20} & 0.37 & $43.27$ & [1] & $0.040$ & $0.16$ & 0.19 & 32.39 & $-4.56$ & 8.0--9.2 \\
 & & & & &  & $8.3\times 10^{-5}$ & $3.3\times 10^{-4}$ &  &  &  &\\
 0321$+$340 & 20 & {50} & 4.89 & $44.76$ & [2] & $7.8\times 10^{-3}$ & 0.03 & 0.04 & 34.26 & $-6.29$ & 9.1--10.3 \\
  && && & & $3.9\times 10^{-4}$ & $1.6\times 10^{-3}$ & & & & \\
 0415$+$379 & 20 & {50} & 3.12 & $45.78$ & [2] & \ldots & \ldots & 0.10 & 34.57 & $-4.89$ & 8.9--10.2\\
  && & && & $5.0\times 10^{-3}$ & $0.02$ & & & & \\
 0430$+$052 & 8 & {40} & 1.14 & $44.94$ & [1] & $9.4\times 10^{-3}$ & $0.04$ & 0.16 & 33.80 & $-4.80$ & 8.6--9.8 \\
  && & &&  & $2.3\times 10^{-3}$ & $0.01$ &  &  &  &\\
 1133$+$704 & 1 & {50} & 2.11 & $44.48$ & [2] & $7.4\times 10^{-3}$ & $0.03$ & 0.13 & 33.75 & $-5.84$ & 8.7--10.0\\
 1228$+$126$^a$ & 10& {20}  & 9.00 & $43.59$ & \ldots & 0.037 & 0.30 & 0.15 & 33.43 & $-7.91$ &\\
 1514$+$004 & 1 & 50 & 1.43 & $45.22$ & [4] & \ldots & \ldots & 0.09 & 33.95 & $-4.77$ & 8.6--9.8 \\
 1637$+$826 & 1 & {40} & 0.75 & $44.80$ & [3] & $0.076$ & $0.30$ & 0.16 & 33.46 & $-4.44$ & 8.3--9.5 \\
 1807$+$698 & 1 & {50} & 0.11 & $45.11$ & [2] & $0.029$ & $0.12$ & 0.08 & 33.77 & $-4.61$ & 8.4--9.7\\
   &&&  &&   & $1.3\times 10^{-3}$ & $5.0\times 10^{-3}$ &  &  &  &\\
 2200$+$420 & 11 & {50} & 4.01 & $45.06$ & [2] & $4.1\times 10^{-3}$ & $0.02$ & 0.15 & 34.32 & $-5.82$ & 9.0--10.3\\
\hline
\end{tabular}
\begin{flushleft}
$^a$ The properties for 1228+126 (M87) were calculated differently than for the other sources. This is because the outer pressure was measured for M87. We assume the mass from the velocity dispersion method. The known mass and pressure along with the Michel's magnetization parameter and the observed position of the break allows reconstructing all the other parameters, whereas for the rest of the sources we need to make an assumption on the total jet power to fully model their properties. The reported maximum Lorentz factor $\gamma\sim 10$ is adopted from \citet{BSM99}.
\end{flushleft}
\end{table*}

In order to probe the lower spin boundary for the other sources with the measured $k$-indices, we propose to use the core widths as upper boundaries for the 
jet width at the break. We use the median core size derived from the structure model fitting of the source brightness distribution performed in the spatial frequency domain at all available epochs at 15~GHz \citep{MOJAVE_XVII}, excluding those cases when the core was modelled as a delta function. This approach is more accurate, as the core size estimates obtained from the stacked maps are subject to overestimation due potential blurring caused by limited accuracy of the core position used to align the images in a stacking procedure. These widths are taken $0.5$~mas  to the left of the horizontal line in \autoref{f:k_vs_rs} \citep[see details in][]{Kovalev20_r1}. This means that the core widths may be at the conical domain, yielding a wider jet. The observed core may fall into the presumed parabolic domain. In this case its position must be close to the shape transition point. Thus   
the core width may reflect the true jet width at the break.

In \autoref{spins_rm} for all 39 sources in our sample we plot the spin range for $\sigma_{\rm M}\in[5;\;50]$ against the redshift $z$. We use the mass estimate method basing on the BLR size-luminosity relation.
We calculate the mean and median spin values for $\sigma_{\rm M}=50$ basing on typical magnetization for the sources with detected recollimation. Both mean and median for the spin lower boundary are equal to $0.14$ and $0.09$, correspondingly. Thus the lower spin estimates presented in \autoref{spins_rm} are in good agreement with the expected from the evolution modelling spin values of the order of $0.1$. We do not see the rise of spins to extreme values $>0.9$ predicted for $z>2$, may be due to a scarce number (four) of sources with such redshifts.

We also observe that spins obtained from the core width have roughly the same values of the spins obtained directly from the jet width at the break. This may be explained if the core widths for this sample are good approximations to the jet break widths. The cores are taken at the distance 5 mas to the left of the leftmost end of a horizontal line which represents the behaviour of $k$-index as a function of a distance along the jet in \autoref{f:k_vs_rs}. As the core widths seem to correspond to the jet widths at the GTR, we conclude that the leftmost line ends can approximate the real position of a shape transition for these sources. This supports the prediction of the position of a jet GTR at $r_\mathrm{break}\in(10^5;\;10^6)r_\mathrm{g}$ \citep{Kovalev20_r1}.

We see a very tentative upward trend of the lower spin estimates in \autoref{spins_rm}. For the upper ends of these lower limits in spin we find the slope of $\log_{10}|a_*|$ as a function of $\log_{10}(1+z)$ being $0.84$, which is close to the result by \citet{Daly-11} for smaller redshifts. However, this correlation may be partly artificial since we have lower estimates rather than determined values and due to our redshift-dependent linear resolution.

Here we should note that the results for spins depend on the chosen jet model. We set the integrals of motion \citep[see details in][]{BCKN-17, Kovalev20_r1}, 
which are associated with an outflow from the vicinity of a black hole.
No extended disc flow with self-similar integrals, as in the Blandford--Payne model, was used in this work. Adding this outflow may affect the result in the following way: the expected jet radius at the break will be larger for the same light cylinder radius. This may boost the values for a BH spin $|a_*|$, putting all the sources in the expected spin range. We plan to address this issue in a forthcoming publication. 

Another reason for observing low spin values may be a selection effect. The jet width at the break point depends linearly on a light cylinder radius. Thus all the other physical parameters being equal, smaller $|a_*|$ correspond to a wider jet, with the GTR more easily resolved by observations.

\section{Black hole masses}
\label{ss:mbhdiff}

Another possible explanation of the obtained low spin may be uncertainty in SMBH masses. Changes in the assumed values of the BH mass strongly affect the spin values. For 1H~0323$+$342 there is an argument that the BLR size-luminosity relation method may underestimate the BH mass \citep{Hada18, LeonTavares14}, and this can also be the case for other sources. Multiple studies of flaring events in AGNs \citep{Arshakian-10, LeonTavares-10, LeonTavares-13, Chavushyan-20} present evidence that this activity results in broad emission line fluctuations, affecting the possibility of using the correlation BLR motion--luminosity in emission lines to estimate the BH masses.

\begin{figure}
\centering
\includegraphics[width=\columnwidth, trim=0cm 1cm 0cm 0cm]{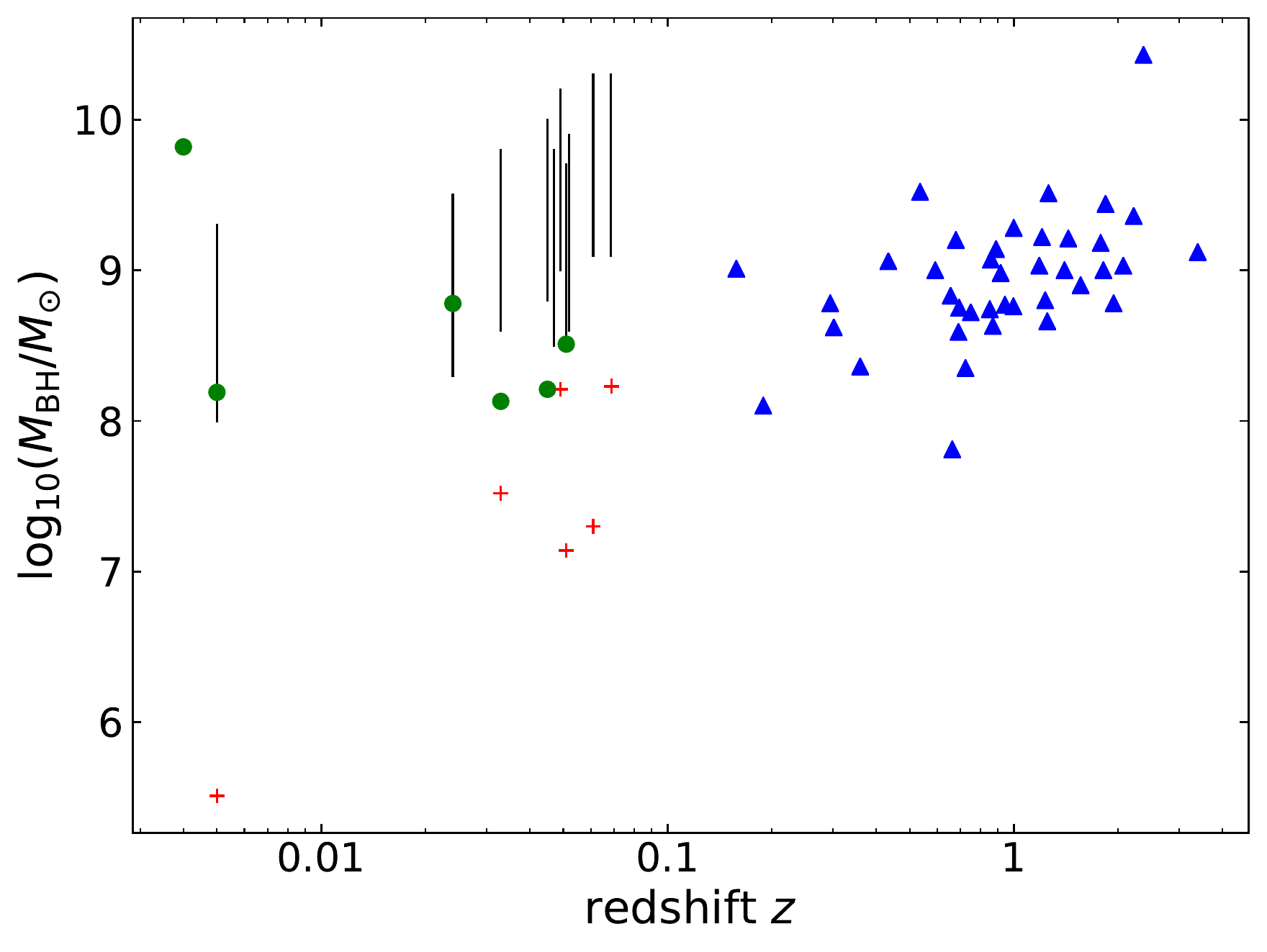}
\caption{Distribution of the BH masses as a function of a redshift. The filled blue triangles designate the sources with the masses determined by the correlation between the BLR size and the luminosity method. The red crosses correspond to the sources with the observed break in the jet shape. The filled green circles represent the sources with the detected jet shape break and the mass estimate by the velocity dispersion method. The vertical lines designate a range for the mass for $a_*\in[0.1,\;0.99]$ for the sources in Table~2. The upper end corresponds to $a=0.99$, and the lower end corresponds to $a=0.1$.}
\label{Mofz}
\end{figure}

\begin{figure}
\centering
\includegraphics[width=\columnwidth, trim=0cm 1cm 0cm 0cm]{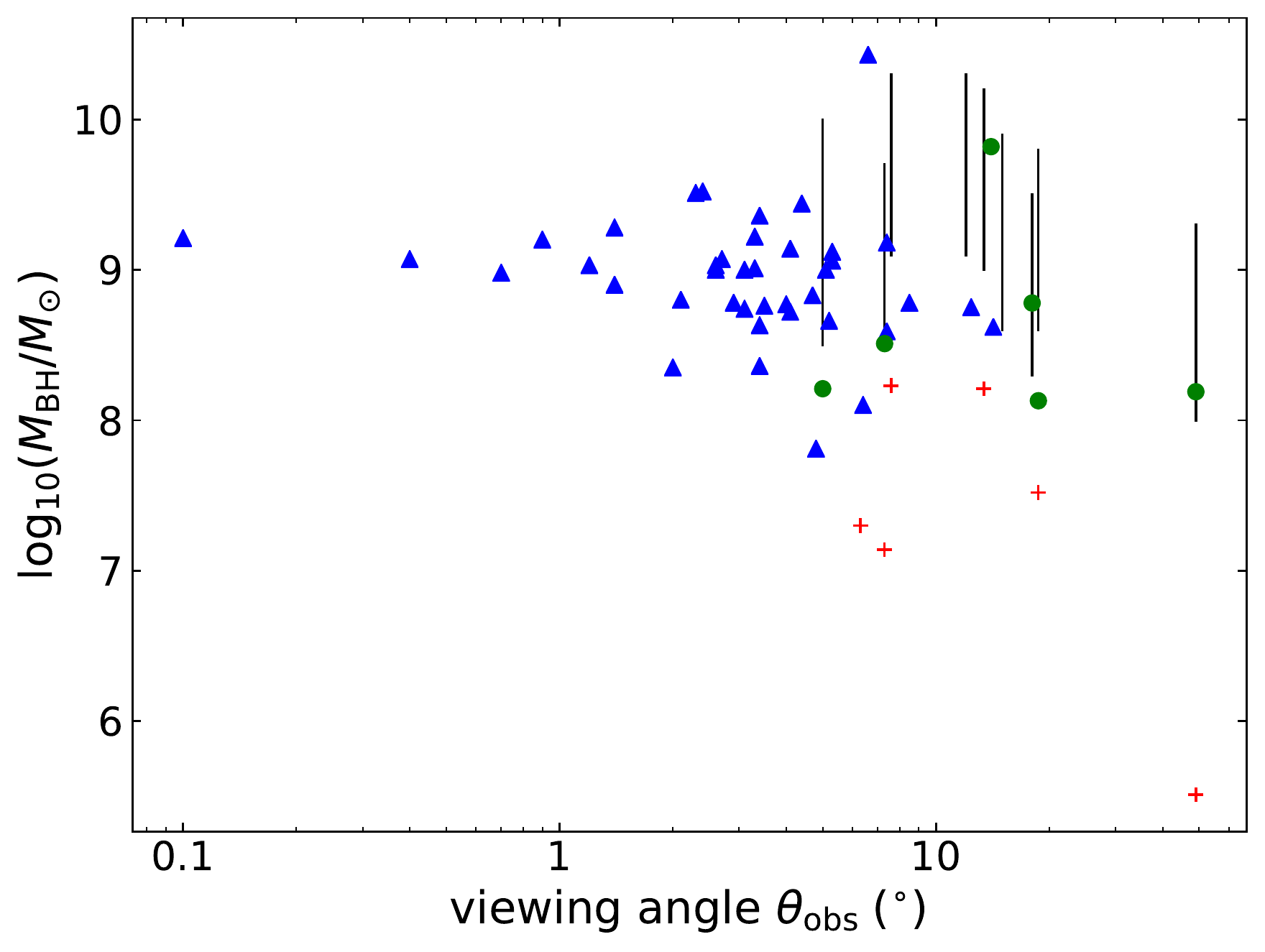}
\caption{Estimated black hole mass as a function of a viewing angle $\theta_{\rm obs}$ for our sample. Designation is the same as in \autoref{Mofz}.}
\label{Mofth}
\end{figure}

We map the BH mass obtained by different methods, as a function of a redshift $z$ (see \autoref{Mofz}). For the small redshifts, the $M_{2}$ values display a larger dispersion than for larger $z$, while all the $M_{1}$ values cluster around $10^9\;{\rm M_{\odot}}$. All the sources with the detected jet shape break have small $z$.
We plot also the BH masses as a function of a jet viewing angle (\autoref{Mofth}).
Larger viewing angles $\theta_{\rm obs}$ result in smaller $M_2$ values; this trend is not observed for $M_1$. 
In order to check whether there is a significant anticorrelation between $M_{2}$ and $\theta_\mathrm{obs}$, we chose the sources with the masses found by the correlation between the size of the BLR and the luminosity method and excluded the source 0238$-$084 (the lower right point in \autoref{Mofth}), as a possible strong driver of the correlation. For this sample of 44 sources we obtained the Spearman's rank correlation coefficient between $M_{2}$ and $\theta_\mathrm{obs}$ to be equal to $-0.44$, with the p-value (chance correlation) equal to $0.0026$. Thus we obtained the significant anti-correlation of these values. 

We have checked that the obtained anti-correlation $M_2 - \theta_{\rm obs}$ cannot be explained by a physical
effect. Indeed, in this case, this would imply that black holes with smaller masses launch more powerful jets, which we can detect for larger viewing angles with weaker relativistic boosting. The jet power may be estimated as $W_j\propto c(\Psi_0/\pi R_{\rm L})^2$ \citep{Beskin10}. The smaller mass with the same BH spin rate leads to a smaller light cylinder radius and a greater power. The total magnetic flux relates to the mass accretion rate as $\Psi_0\propto\sqrt{\dot{M}}r_g$, For the Bondi accretion $\dot{M}\propto M^2$ we have a rough relation $W_j\propto M^2$ --- the well known relation for the Blandford--Znajek process \citep{BZ-77, Moderski96, Daly-19}.

Thus we reject the relation ``the more powerful jet --- the smaller SMBH mass'' and conclude that 
the obtained anti-correlation may be caused by a possible bias in the mass estimate method itself. 

The sample of $M_{1}$ does not have any significant anti-correlation with the observation angle, although this is based on the sample of seven sources only. 

The obtained anti-correlation $M_2 - \theta_{\rm obs}$ suggests that it is important to take into account the observation angle while using the BLR size and luminosity correlation method to find a BH mass. A large enough viewing angle may change the geometry factor or, due to partial veiling of an inner region (by a torus, for instance), affect the observed spectral luminosity or the lines width.
This effect is important for the closer sources, because we can detect the source with a big enough viewing angle and, hence, without sufficient Doppler boosting, only at the smallest distances. The further and fainter sources can be detected only with a small viewing angle due to Doppler beaming. Thus we suggest that the masses $M_2$ of the sources with the detected break in a jet shape, obtained by the BLR size--line luminosity correlation method, may be underestimated. This is also relevant for the source 0321$+$340. 

We propose an alternative method of the BH mass estimate from the observations of a jet GTR. 
Our modelling
does not provide explicitly the BH mass or spin. As being based on MHD modelling, it provides the light cylinder radius $R_{\rm L}$ that can be related to both these values by \autoref{astar}.
The light cylinder radius may 
be readily found using \autoref{RL}, and the BH mass may be obtained by fixing the BH spin $|a_*|$ in the range expected for active galaxies. 

Using our model and the BH spin interval $|a_*|\in(0.1,\,0.99)$, we find the BH masses for 10 sources with the observed jet shape transition (see \autoref{t:break}). In \autoref{tableParam} we present the results for the BH masses, and we plot the mass ranges as function of redshift and jet viewing angle in \autoref{Mofz} and \autoref{Mofth}, respectively. 
We see that the predicted BH masses are typically slightly larger than those obtained by the velocity dispersion measurements and are much larger than those obtained by other methods. Our mass estimates for three sources matches the masses obtained by the kinematics method. We may add to this positive result the mass estimate for M~87 from \citet{nokhrina2019}. We note that our result for 0321$+$340 (1H~0323$+$342) is very close to the mass obtained by \citet{LeonTavares14}. We also obtained the yet unknown masses for 0111$+$021 and 1514$+$004.
We note that our mass estimates fall much better into the general trend for masses for AGNs than the values obtained by the BLR-size--line luminosity correlation for the large observational angles.

\section{Ambient pressure}
\label{ss:BreakPos}

\begin{figure}
\centering
\includegraphics[width=\columnwidth, trim=0cm 1cm 0cm 0cm]{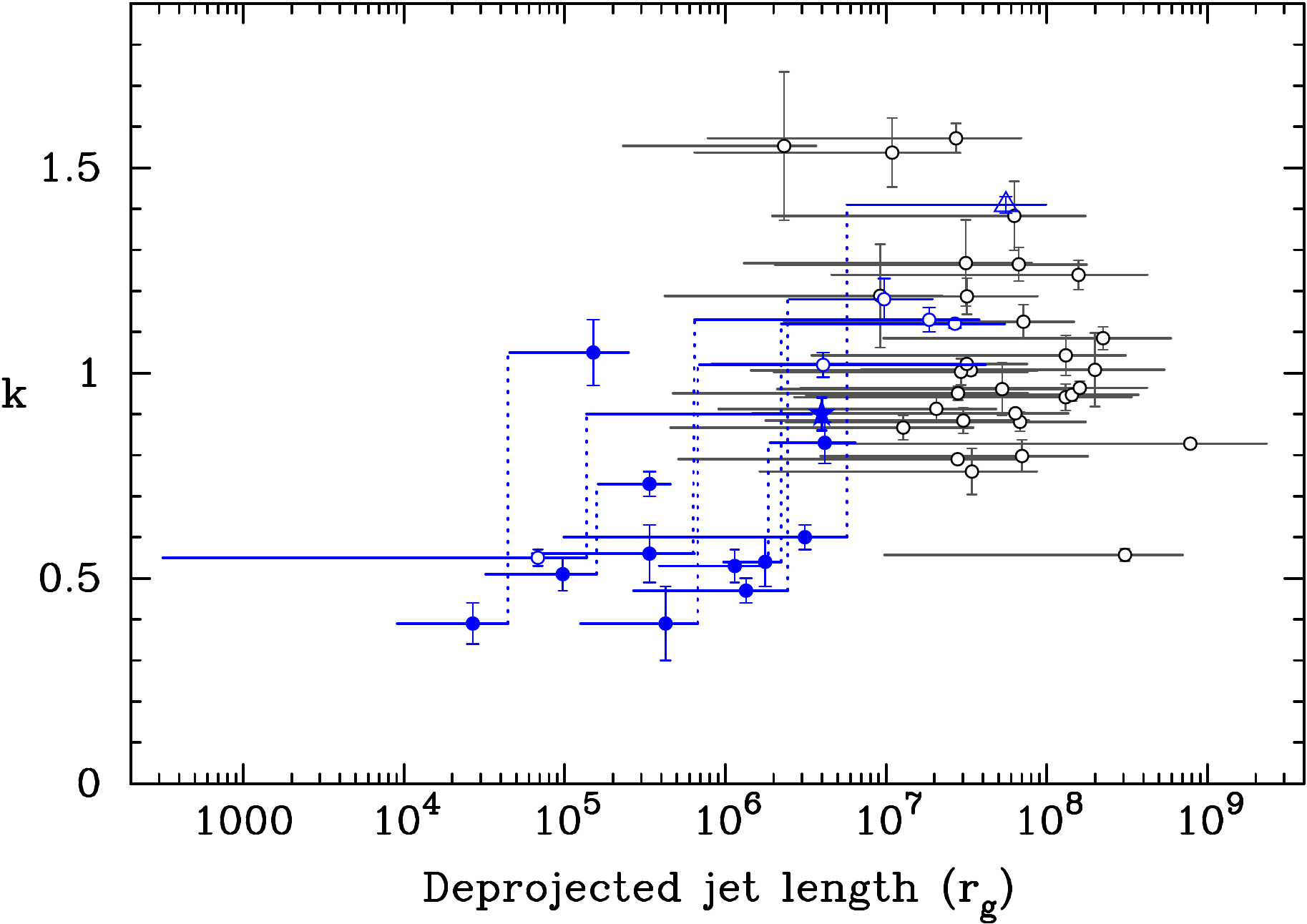}
\caption{Best fit $k$-index values against the deprojected distance measured in the gravitational radius from 15~GHz VLBA core for the sources listed in Table~1 \citep{Kovalev20_r1} with the measured redshift and the observing angle. 
The filled dots show fits at 15~GHz only, while the empty circles denote the results from analysing measurements at 15 and 1.4~GHz. 
The horizontal lines denote the scale at which the $k$-index was measured for every target. 
The symbols are placed at the median distance of the analysed jet portion. Nine AGN with the detected jet shape transition are shown in blue: 0238$-$084, 0321+340, 0415+379, 0430+052, 1133+704, 1637+826, 1807+698, 2200+420 and M87. The data for 0321+340 and M87 are taken from \citet{Hada18} and \citet{nokhrina2019}, respectively.
The black hole masses are shown in Table~1 \citep{Kovalev20_r1}. If available, we use the estimates based on the velocity dispersion method, otherwise --- those from the BLR-size--luminosity correlation technique.
In contrast with \citet{Kovalev20_r1}, we assume the mass for 0321$+$340 obtained by \citet{LeonTavares14}.
} 
\label{f:k_vs_rs}
\end{figure}

Using the position $r_{\rm break}$ of a GTR along the jet is not so straightforward as using the jet width $d_{\rm break}$. 
First, let us introduce an important conserved value.
The observed parabolic jet shape $d\propto r^{0.5}$ upstream the break point together with an assumption of an equilibrium of the jet and the ambient pressure $P\propto r^{-2}$ provide the jet pressure at the boundary is equal to $P_{\rm jet}\propto d^{-4}$, independently of our model. We assume a pressure balance $P=P_{\rm jet}$ at the jet boundary. So, in the acceleration and collimation zone, the relation $P_{\rm jet}d^4$ must be approximately constant.
Within our model we confirm the relation $P_{\rm jet}\propto d^{-3.7}\approx d^{-4}$ \citep{BCKN-17, Kovalev20_r1}.
At the jet base on the scales of a few gravitational radii the poloidal magnetic field may be considered uniform, and the pressure is due to magnetic pressure $P\propto B^2$. On this scale the conserved value $Pd^4$ is proportional to the square of a total magnetic flux $\Psi_0^2$. 
So, the conservation of $Pd^4$ may be written as
\begin{equation}
\frac{P_{\rm break}d_{\rm break}^4}{\Psi_0^2}\approx{\rm const}.
\label{const1}
\end{equation}
Using now \autoref{P_star} and \autoref{Pbrdim},
we rewrite it as
\begin{equation}
{\rm const}\approx\frac{P_{\rm break}d_{\rm break}^4}{(2\pi\sigma_{\rm M})^2p_0R_{\rm L}^4}=\frac{4}{\pi}\frac{P_*d_*^4}{\sigma_{\rm M}^2}.
\label{Pd4}
\end{equation}
The left hand side of \autoref{const1} is constant and does not depend on the initial jet magnetization, so the right hand side of \autoref{Pd4} does not depend on it, either. We check it in our semi-analytical modelling and obtain, indeed, that $\sigma_{\rm M}^2/(P_*d_*^4)\approx 1.60\pm 0.09$.

This result means that the jet shape transition and, consequently, the acceleration pattern along the jet depend on the ambient pressure and occurs when the jet attains a certain width. In order to fix $r_{\rm break}$, we need to know $P_0$ at some distance $r_0$. Conversely, for the known $\Psi_0$ and the measured $d_{\rm break}$ we can estimate pressure at $r_{\rm break}$.

Bearing in mind this observation, it would be ideal to plot $r_{\rm break}$ in units of the Bondi radius $r_{\rm B}\sim GM/c_s^2$, which is a natural length unit characterising ambient pressure. It depends on the gravitational radius and the sound speed $c_s$ at the sonic radius. The estimation of a Bondi radius needs complicated measurements of the ISM temperature and density, and is known for a few sources. To the best of our knowledge, in our sample of 11 sources with a detected GTR, $r_{\rm B}$ was measured only for M~87 by \citet{RF15} having a value $\lesssim 10^6\;r_{\rm g}$. Since the Bondi radius is expected to be in the interval $10^5-10^6\;r_{\rm g}$ \citep{BMR-19}, $10^6\,r_{\rm g}$ may be useful as a proxy for $r_{\rm B}$. 
Thus we plot the observed $k$-index values as a function of the de-projected distance along jets in physical units of a gravitational radius. By doing this, we expect to find the jet shape break positions $r_\mathrm{break}$ to be bounded, which provides an instrument for constraining typical $P_0$. The result is presented in \autoref{f:k_vs_rs}. We observe that all the sources with the detected jet shape break have $r_\mathrm{break}/r_\mathrm{g}\in(10^5,\,10^6)$. If we assume that the jet shape break is a common phenomenon, then most of the sources with only a conical part observed demonstrate that the jet shape break position is expected to satisfy roughly $r_\mathrm{break}/r_\mathrm{g}<10^6$. Thus we may conclude that the expected jet shape break will be observed mostly in the same interval. This result is independent of the possible mass underestimation discussed above: there is only one source in \autoref{f:k_vs_rs} that is plotted in units of $r_g$, determined by the $M_{2}$ (0430+052) --- the second rightmost source. If, indeed, its mass is higher, the jet GTR will move to smaller distances, falling better into the interval $r_\mathrm{break}/r_\mathrm{g}\in(10^5,\,10^6)$. 

We propose to decouple the total magnetic flux $\Psi_0$ and the ambient pressure amplitude $P_0$ in \autoref{rbreak} as follows. The total power of an initially magnetically dominated outflow relates to the total magnetic flux in a jet by the expression \citep{BZ-77, Moderski96}
\begin{equation}
W_{\rm j}=\frac{c}{32}B_{\rm g}^2r_{\rm g}^2\left(\frac{a_*}{1-\sqrt{1-a*^2}}\right)^2.
\label{Power-BZ}
\end{equation}
The expression coincides with the estimate by
\citet{Beskin10}
\begin{equation}
W_{\rm j}=\alpha c\left(\frac{\Psi_0}{\pi R_{\rm L}}\right)^2\label{Pjet}
\end{equation}
with the numerical coefficient $\alpha=1/8$. In general, this coefficient depends on the energy integral $E(\Psi)$. Is was shown by \citet{N17Fr,Nokhrina18} that the average jet power relates to the 
initial electromagnetic power above with the coefficient $\alpha=1/8$.
The average jet power can be correlated to the jet luminosity at radio frequencies in 200$-$400 MHz range \citep{Cav-10} from the CATS database \citep{CATS05}. This allows us to estimate the total magnetic flux in the jet using \autoref{Pjet}, and the ambient pressure amplitude at the break (or at any other distance inside the Bondi sphere) $P_{\rm break}$ using \autoref{const1} and \autoref{Pd4}.

\subsection{Sources with the break}
\label{ss:wbreak}

For the 11 sources with the observed break we can readily constrain the pressure $P_{\rm break}$ at the observed jet break point. The result is presented in the 11th column of \autoref{tableParam}. 

The direct comparison, although on the scales much greater that the break point, is possible for the source 1637$+$826: the measured ambient pressure profile by \citet{Evans-05} provides $P\approx (2- 4)\times 10^{-10}\;{\rm dyn/cm^2}$ at the distance $\sim 480$~pc. We estimate the pressure predicted by our measurements and the model using the Bondi pressure profile $P\propto r^{-2}$, and obtain $P\approx 19\times 10^{-10}\;{\rm dyn/cm^2}$ at the same distance of $480$~pc. This value is less than order of magnitude larger than the direct measurements value. It is a good correspondence, given the possible errors in the jet GTR determination, uncertainties in direct temperature and density measurements, uncertainties in the model and possible deviation of pressure from the Bondi profile at large distances.

There is more data for the M~87 jet than for the other sources in our sample, so we can check the predicted by our model mass accretion rate against observations using a Faraday rotation \citep{Kuo-14}. 
For the M~87 jet \citet{RF15} measured a temperature and a particle number close to the expected Bondi radius. This data allows us to estimate the pressure $P_0=4.5\times 10^{-10}\;{\rm dyn\;cm^{-2}}$ at $r_0=0.22$~kpc. Using this data for estimating $P_{\rm break}$ using \autoref{Pscale}, we constrain the total magnetic flux $\Psi_0\approx 3\times 10^{33}\;{\rm G\;cm^2}$ using \autoref{Pd4}. The relation between a total magnetic flux and a mass accretion rate
\begin{equation}
\Psi_0=\phi r_{\rm g}\sqrt{\dot{M}c}
\label{disk}
\end{equation}
depends on a dimensionless flux $\phi$, which assumes values from the order of unity for a standard and normal evolution disc (SANE) up to $~50$ for the magnetically arrested disc (MAD) \citep{Tchekhovskoy_11, Narayan-12}.
\citet{nokhrina2019} used a Bondi accretion rate $\dot{M}=0.1\,{\rm M_{\odot}/yr}$ \citep{DiMat-03} using $\phi\approx 3$. This is in contrast with the results by \citet{EHT_V} who excluded a low spin for a SANE disc.
Using our estimate for a total magnetic flux and setting $\phi=50$, corresponding to MAD, we constrain the mass accretion rate for M~87 as $\dot{M}=1.8\times 10^{-3} \,{\rm M_{\odot}/yr}$, which is in agreement with the accretion rate estimate $9.2\times 10^{-4}\,{\rm M_{\odot}}$ based on Faraday rotation measured by \citet{Kuo-14}.

\begin{figure}
\centering
\includegraphics[width=\columnwidth, trim=0cm 0.8cm 0cm 0cm]{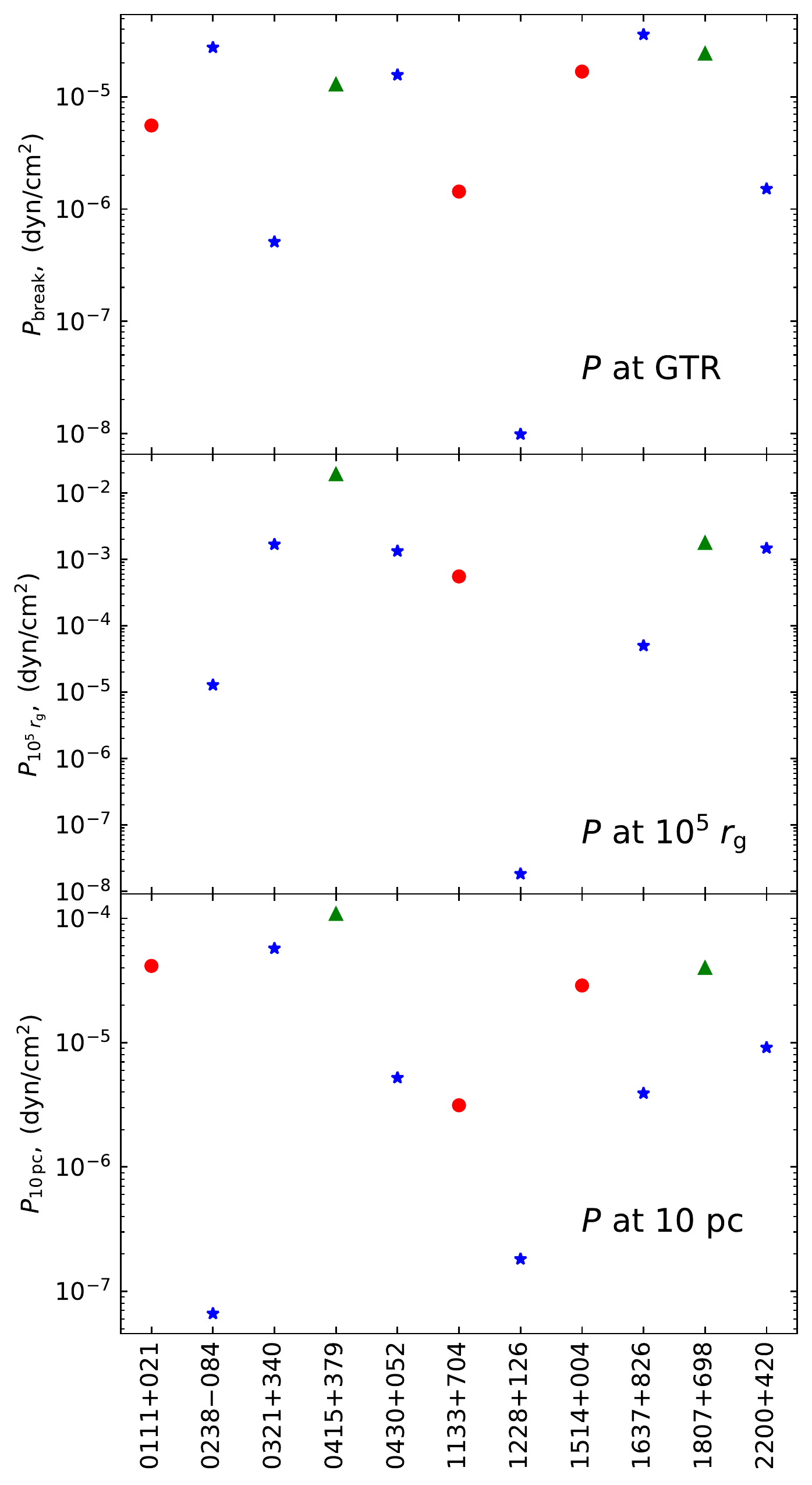}
\caption{Pressure at different points for 11 sources listed in \autoref{t:break}. For the second plot sources 1 and 8 are absent due to unknown mass. Fanaroff--Riley class is highlighted by marker: FRI is a blue star, FRII is a green triangle and unknown class is a red circle.}
\label{Fig_pressure}
\end{figure}

The pressure values estimated within our model and under assumption of the Bondi accretion, given at different distances, provides us with a very different kind of information. As we discussed at the beginning of \autoref{ss:BreakPos}, the pressure at the GTR $P_{\rm break}$ reflects the jet and black hole properties only. It equals to the jet inner pressure at the boundary at the point of acceleration saturation (roughly $\sigma=1$), while this value depends on the light cylinder radius and the initial jet magnetization.
We see in the upper panel of \autoref{Fig_pressure} that the scatter in $P_{\break}$ spans less than two orders of magnitude (with the exception of M~87), reflecting, probably, the alike jet inner properties of these sources.

Since the expected Bondi radius should be in the range $(10^5-10^6)\;r_{\rm g}$, one can easily extrapolate the estimate on pressure at GTR to the pressure $P_{5}$ at a fixed distance $10^5\,r_{\rm g}$. In this case we step away from the jet inner properties and probe pressure depending on the gravitational radius value. This pressure reflects both the ambient medium and the central source through its mass. The dispersion in the values of $P_5$ (with exception of M87) is much larger than in the previous case, spanning three orders of magnitude. 

We can also employ the third pressure value at some fixed distance close enough to $10^5\,r_{\rm g}$, to ensure being inside the Bondi sphere. For the typical BH masses the distance $r=10$~pc roughly corresponds to 
$10^5\,r_{\rm g}$ and does not depend on a black hole mass. The pressure value at this distance should reflect the ambient medium within the Bondi sphere only. We observe that the overall scatter in pressure values is three orders of magnitude. We can also see, although very tentatively, the greater pressure for the Fanaroff-Riley class II sources. There are presumably two outliers (radio galaxies 0238$-$084 and M87). Without them, the scatter in values of $P_{10\,{\rm pc}}$ is around 30 pointing to the uniform ambient medium properties for these sources. 

There is no dependence of the results discussed above on the choice of an exponent $b$ in the Bondi pressure profile \autoref{Pscale}. This value affects only the predicted $k$-indices in modelled jets. We should mention that we cannot reproduce the observed $k$-indices in a jet form for the full set of the sources with the detected jet shape break. We reproduce well the $k$-indices (within the errors) for four sources listed in \autoref{t:small}. The data for M~87 is from \citet{nokhrina2019}.
This is because within our semi-analytical model the difference in the modelled powers $k_1$ in parabolic and $k_2$ in conical domains is less than $0.5$. We think that a different choice of the jet temperature and the sound velocity will affect $k$-indices in the jet boundary shape for $b\approx 2$. We will address this issue in a forthcoming publication. 

\begin{table}
\caption{Modelled $k$-indices for the sources with a good correspondence to the the observed ones.
(1) the source name (B1950);
(2) the power in the pressure dependence law; 
(3) the $k_1$ index (parabolic); (4) the $k_2$ index (conical). 
\label{t:small}}
\centering
\begin{tabular}{cccc}
  \hline\hline
 Source & b & $k_1$ & $k_2$ \\
 (1) & (2) & (3) & (4) \\ 
\hline
1133$+$704 & 2.00 & 0.550 & 0.793 \\
1228$+$126 & 2.07 & 0.570 & 0.820 \\
1514$+$040 & 2.25 & 0.612 & 0.864 \\  
1637$+$826 & 1.90 & 0.522 & 0.754 \\
\hline
\end{tabular}
\end{table}

\subsection{Sources with unresolved break point}
\label{withoutbreak}

The obtained jet shape break position in the interval of approximately $10^5-10^6\,r_{\rm g}$ allows us to constrain roughly the pressure of the ambient medium which collimates a jet for the nearby sources. Using \autoref{Pd4}, we obtain
\begin{equation}
P_{5}=0.25\frac{\Psi_0^2}{d_{\rm break}^4}\left(\frac{r_{\rm break}}{10^5\,r_{\rm g}}\right)^2.
\end{equation}
For the typical magnetic flux $\Psi_0\sim 10^{33}\,{\rm G\; cm^2}$ and $d_{\rm break}\sim 1\,{\rm pc}$, we find that at $10^5 r_{\rm g}$ the ambient medium pressure value must be in the interval $10^{-9}-10^{-7}\,{\rm dyn\;cm^{-2}}$. We observe that the typical expected pressure at the Bondi radius is consistent with the measurements for M~87 \citep{RF15}. Due to the different jet inner properties (magnetic flux, light cylinder radius), the real pressure must be scattered around the found above interval.  

\section{Summary}
\label{s:summary}

We used data of the jet width at the geometry transition region and distance to the jet apex to estimate several physical properties of a jet and the central black hole. 

For the sources with the detected jet shape transition we constrain within our model the black hole spin values. We observe the dichotomy in the spin values for the sources with mass estimates made by different methods, with higher spin values corresponding to the velocity dispersion methods. The median value of spins 0.14 for the latter sample is consistent with the lower values $\gtrsim 0.1$ predicted by the spin evolution for the super massive black holes at low redshifts \citep{Barausse12, Volonteri13, Sesana14}. The obtained spin values may have an observational bias, as for the lower spins both the position of the geometry transition region and a jet width at this point are larger, so this change in a shape geometry can be resolved more easily. 

We do not observe any trend ``larger spin -- more powerful jet'' in our results \autoref{tableParam}. The first reason may be connected with the method of an estimation a jet power based on a correlation of a radio flux with a power needed to blow the cavities around jets \citep{Cav-10}. This method reflects the averaged over a large amount of time jet power, which may not correlate with the spin on short-time scale. We do not see a contradiction of a high jet power and the low spin for a source, because the relation between the two, given by \autoref{Power-BZ} and \autoref{Pjet} \citep{BZ-77, Moderski96, Beskin10}, includes the total magnetic flux. The flux estimates based on a spin and a jet power values assumes the values of the order of $10^{32}-10^{34}\;{\rm G\;cm^2}$ in good agreement with the expected flux values \citep{ZCST14, Finke19}.

For each of other 39 sources without a detected jet boundary transition, but with the measured $k$-index close to unity, we use the core width as the upper limit of the jet width at the geometry transition region and obtain the lower limit on the source black hole spin. The median spin value 0.09 for this sample is consistent with the predictions of the spin evolution models \citep{Barausse12, Volonteri13, Sesana14}. It also consistent with the median spin value $0.14$ for the sources with the measured jet width at the geometry transition region with mass estimate by the velocity dispersion method. This means that the core width may be an acceptable approximation for the jet width at the geometry transition region. This fact supports the prediction that the geometry transition occurs at a distance of $(10^5,\,10^6)$ in units of a gravitational radius $r_{\rm g}$.

We find that the black hole masses obtained by the relation between the BLR size and line luminosity have a significant anti-correlation with the viewing angle. We propose that there may be a bias in this method, related either to a geometry factor or to partial veiling of the molecular torus for large enough observational angles, which may lead to a mass underestimate. We propose a new mass estimate method, based on the relation between the jet width at the geometry transition region and the light cylinder radius. This method allows us to find a BH mass range assuming that the spin range is $|a_*|\in(0.1,\,0.99)$ for active galaxies. 
The lower limit of the mass constraint based on our method is in agreement with the mass obtained using velocity dispersion method (having an accuracy of within factors of a few, estimated by \citet{WU02}).
The proposed method may be useful for the BH mass estimates for the sources with moderate redshifts: those, for which the deprojected distances of the order of $(10^5,\,10^6)\;r_{\rm g}$ may be resolved by very large baseline interferometry observations. 

The position of a jet transition point along the jet provides an estimate of the value of an ambient pressure amplitude. For the source 1637$+$826 the pressure, estimated from the jet geometry transition region position, is in agreement with the measurements obtained by \citet{Evans-05}. We note that the ambient pressure values at different scales reflect different properties of AGNs. The value of pressure at the geometry transition region $P_{\rm break}$ depends on the jet and SMBH inner properties only. The scatter in values of $P_{\rm break}$ provides us the cumulative scatter in the total magnetic flux, the BH spin and the initial magnetization of the regarded sources. Assuming that for all our 11 sources the distance of 10~pc is inside or close to the Bondi radius, by calculating pressure at this distance, we probe the properties of the ambient pressure exclusively.

The results presented here are obtained for a small sample of AGN. 
We have recently started a dedicated VLBA observing program for several dozens of nearby active galaxies to perform a systematic search and study of the jet shape transition. This will provide a robust set of data to test our findings and enhance conclusions on the properties of SMBHs, jets, and conditions in the ambient medium.  

\section*{Acknowledgements}

We thank the anonymous referee for suggestions that helped to
improve the paper. We thank Eduardo Ros, Vasily Beskin, Matt Lister, Andrei Lobanov, Alexander Plavin, Lev Titarchuk, and Tigran Arshakyan for the valuable comments and suggestions.
This study has been supported
by the Russian Science Foundation: project 20-62-46021.
This research made use of the data from the MOJAVE database\footnote{\url{http://www.physics.purdue.edu/MOJAVE/}} which is maintained by the MOJAVE team \citep{MOJAVE_XV} and the data accumulated by the CATS data base \citep{CATS05}.
This study makes use of 43 GHz VLBA data from the VLBA Boston University Blazar Monitoring Program\footnote{\url{http://www.bu.edu/blazars/VLBAproject.html}} funded by NASA through the Fermi Guest Investigator Program.
The VLBA is an instrument of the National Radio Astronomy Observatory. The National Radio Astronomy Observatory is a facility of the National Science Foundation operated by Associated Universities, Inc. 
This research made use of NASA's Astrophysics Data System.

\section*{Data availability}
The data underlying this article is available in \citet{Kovalev20_r1}, the original datasets were derived from sources in the public domain at http://www.physics.purdue.edu/astro/MOJAVE/\\allsources.html.

\bibliographystyle{mnras}
\bibliography{yyk}

\bsp    
\label{lastpage}
\end{document}